\documentclass[apj]{emulateapj}

\usepackage{graphicx,array,booktabs}

\shorttitle{Large Scale Asymmetries in Transitional Disks}
\shortauthors{P\'erez et al.}

\newcommand{\SR}{SR~21}
\newcommand{\SAO}{SAO~206462}
\newcommand{\OphIRS}{Oph~IRS~48}
\newcommand{\HD}{HD~142527}
\newcommand{\LkHa}{LkH$\alpha$~330}
\begin{document}

\title{Large-Scale Asymmetries in the Transitional Disks of SAO~206462 and SR~21}

\author{Laura M. P\'erez\altaffilmark{1,2},
Andrea Isella\altaffilmark{3}, 
John M. Carpenter\altaffilmark{3},
Claire J. Chandler\altaffilmark{1}
}
\altaffiltext{1}{National Radio Astronomy Observatory, P.O. Box O, Socorro NM 87801, USA}
\altaffiltext{2}{Jansky Fellow}
\altaffiltext{3}{California Institute of Technology, 1200 East California Blvd, Pasadena, CA 91125, USA}

\begin{abstract}

\noindent 

We present Atacama Large Millimeter/submillimeter Array (ALMA) observations in the dust continuum (690~GHz, 0.45~mm) and $^{12}$CO~$J=6-5$ spectral line emission, of the transitional disks surrounding the stars \SAO\ and \SR.
These ALMA observations resolve the dust-depleted disk cavities and extended gaseous disks, revealing large-scale asymmetries in the dust emission of both disks.
We modeled these disks structures with a ring and an azimuthal gaussian, where the azimuthal gaussian is motivated by the steady-state vortex solution from \citet{2013Lyra}. 
Compared to recent observations of \HD, \OphIRS, and \LkHa, these are low-contrast ($\lesssim2$) asymmetries. 
Nevertheless, a ring alone is not a good fit, and the addition of a vortex prescription describes these data much better. 
The asymmetric component encompasses $15\%$ and $28\%$ of the total disk emission in \SAO\ and \SR\, respectively, which corresponds to a lower limit of $2M_{Jup}$ of material within the asymmetry for both disks.
Although the contrast in the dust asymmetry is low, we find that the turbulent velocity inside it must be large ($\sim20\%$ of the sound speed) in order to drive these azimuthally wide and radially narrow vortex-like structures.
We obtain residuals from the ring and vortex fitting that are still significant, tracing non-axisymmetric emission in both disks. 
We compared these submillimeter observations with recently published H-band scattered light observations.
For \SR\, the scattered light emission is distributed quite differently from submillimeter continuum emission, 
while for \SAO\, the submillimeter residuals are suggestive of spiral-like structure similar to the near-IR emission.

\end{abstract}


\section{Introduction}
The structure of a circumstellar disk is expected to dramatically change during the planet formation process. In a few~million years an optically thick massive protoplanetary disk will transform into an optically thin debris disk with little material \citep[e.g.][]{2007ApJ...662.1067H}.
Processes like viscous accretion, photo evaporation, grain growth, and planet formation shape the structure of circumstellar disks \citep{2001MNRAS.328..485C,2005A&A...434..971D,2008ApJ...678L..59I}, and in particular, the process of planet formation creates gaps, cavities, and asymmetries that can be directly observed \citep{2005ApJ...619.1114W}.
Circumstellar disks with gaps and/or cavities lack warm dust close to the star while still possessing a significant reservoir of cold dust in the outer disk. Such disks are classified as transitional disks, and plenty is known about their inner disk gaps/cavities \citep[e.g.][]{2008A&A...490L..15D,2009ApJ...699..330S,2010ApJ...717..441E,2012ApJ...747..136I}.
However, it is only recently that observations have reached the sensitivity and angular resolution required to study asymmetrical features in the outer disk structure. 
Particularly, the only known examples of high contrast asymmetries in transitional disks at submillimeter wavelengths are 
 \HD\ \citep{2013Natur.493..191C,2013PASJ...65L..14F}, \OphIRS\ \citep{2013Sci...340.1199V}, and \LkHa\ \citep{2013ApJ...775...30I}. To understand the process of planet formation, long wavelength observations are critical, since they are less affected by optical depth, and hence trace the bulk of the mass surface density.

In this letter, we present ALMA observations at 0.45~mm of two disks surrounding the stars \SAO\ and \SR, which exhibit large-scale asymmetries in their dust continuum emission that were not known before with high significance. 
\SAO\footnote{also HD~135344B.} is an isolated Herbig Ae/Be star (F4Ve, $1.7\:M_{\odot}$), at a distance of 142 pc in the Sco-OB2 association \citep{2011A&A...530A..85M}. 
The disk surrounding \SAO\ is classified as a transitional disk, with a large dust-depleted cavity ($R_{cav}=46$~AU) and a massive outer disk \citep{2011ApJ...732...42A,2009ApJ...704..496B}.
Recently, scattered light observations of the polarized intensity at near-IR wavelengths revealed spiral-like structure in the outer disk of \SAO\ \citep{2012Muto,2013A&A...560A.105G}.
\SR\ is a young G3 star ($2.5\:M_{\odot}$) in the $\rho$-Ophiuchus star forming region at a distance of 119~pc \citep{2008A&A...480..785L}. 
The dust-depleted cavity  \citep[$R_{cav}=36$~AU,][]{2011ApJ...732...42A} in this transition disk \citep{2007ApJ...664L.107B,2009ApJ...704..496B} is not completely devoid of material, as rovibrational CO line observations reveal the presence of inner disk gas \citep{2008ApJ...684.1323P,2013ApJ...770...94B}. Furthermore, recent polarized intensity H-band observations of \SR\ by \citet{2013ApJ...767...10F}  display emission inside the cavity radius, and unlike \SAO, the radial profile of emission is quite smooth.

\begin{figure*}[!t]
\begin{center}
\includegraphics[scale=0.16]{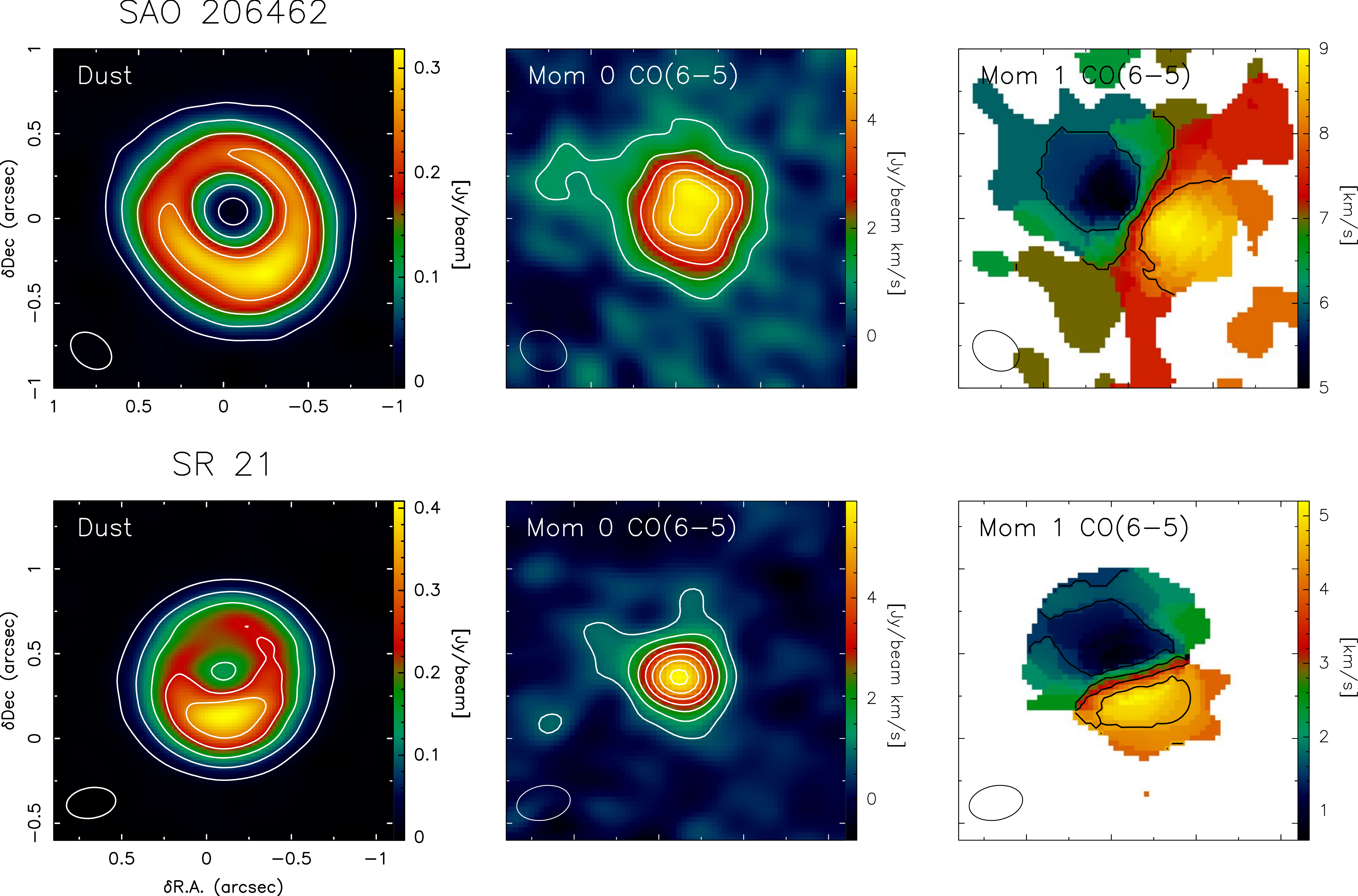}
\caption{ALMA observations of dust and gas emission for the disks surrounding \SAO\ (top row) and \SR\ (bottom row). 
Ellipses indicate beam sizes listed in Table~1.
Left panels: dust continuum emission (color scale) and dust brightness temperature (contours starting at 10~K, spaced by 5~K).
Middle panels: $^{12}$CO~$J=6-5$ moment~0 map, contours start and are spaced by $3\sigma$, where $\sigma$ is the RMS noise level in the map. 
Right panels: $^{12}$CO~$J=6-5$ moment~1 map, contours are spaced by $1\:\textrm{km~s}^{-1}$.}
\label{amp_vs_uvdist}
\end{center}
\end{figure*}

\begin{table*}[]
\vspace{-0.5cm}
\caption{Image properties for dust continuum and spectral line observations.} \label{table_image}
\vspace{-0.3cm}
\begin{center}
\begin{tabular}{l  c c c c c r c c c r}
			&			&			&\multicolumn{4}{c}{Continuum}									& \phantom{a}	& \multicolumn{3}{c}{Spectral line}			\\
\cmidrule{4-7}
\cmidrule{9-11}
Target 		& R.A.$^a$	& Dec$^a$	& $S_{\nu}$$^b$	& RMS 			& Beam		& Beam		&& RMS$^c$	 	& Beam		& Beam	\\
			&  (J2000)		& (J2000)		& (Jy)			& (Jy beam$^{-1}$)& size		& P.A.		&& (Jy beam$^{-1}$)	& size		& P.A.	\\
\midrule
SAO~206462	& 15:15:48.448	& $-$37:09:16.06	&  4.2			&  0.0020		& $0.27'' \times 0.19''$	& 53$^{\circ}$	&& 0.11		&$0.29'' \times 0.22''$	& 60$^{\circ}$	\\

SR~21		& 16:27:10.281	& $-$24:19:12.88	& 3.3				&  0.0011		& $0.29'' \times 0.18''$	& -81$^{\circ}$	&& 0.07		&$0.32'' \times 0.20''$	& -79$^{\circ}$	\\
\bottomrule
\end{tabular}
\end{center}
\vspace{-0.3cm}
$^a$ Phase center from the \citet{2010AJ....139.2440R} catalog, including proper motion (epoch J2000).  \\
$^b$ Flux density integrated inside aperture where emission is above $3\times$RMS noise level. \vspace{0.1cm} \\
$^c$ RMS noise level in a 0.5~km~s$^{-1}$ channel.

\vspace{0.3cm}
\end{table*}

\newpage
\section{Observations and Calibration}

Observations at 0.45~mm (690~GHz) of \SAO\ and \SR\ were obtained during ALMA Cycle~0 between June-July 2012, with baselines ranging from 50 to $500\:$m. 
Four spectral windows were configured to provide 1.875~GHz of bandwidth per polarization, 3840 channels of 0.488~MHz width ($\sim0.2\:\textrm{km~s}^{-1}$), for a total bandwidth of 7.5 GHz in dual polarization,
The bandpass response was calibrated with $3\textrm{C}\:279$ observations, the absolute flux scale was measured using Titan observations, and the phase and amplitude gains were tracked by periodically observing J1427$-$4206 and  J1625$-$2527 for \SAO\ and \SR\ respectively. A total 25~min on-source time was obtained for both targets. 
Visibilities were calibrated and imaged in CASA. Given the high signal-to-noise ratio of these observations amplitude and phase self-calibration was performed after standard phase referencing. 
Table \ref{table_image} lists observational properties of the continuum and spectral line maps, obtained using Briggs weighting with a robust of 0.5.

\begin{table*}[!t]
\caption{Best-fit Results} \label{table_best}
\vspace{-0.3cm}
\begin{center}
\begin{tabular}{  l   r l    r r   c   r r }
			& 					&			&\multicolumn{2}{c}{\SAO}				& \phantom{ab}	& \multicolumn{2}{c}{\SR}	\\
\cmidrule{4-5}
\cmidrule{7-8}
	 		& 					&			& \multicolumn{1}{c}{Ring} 			& \multicolumn{1}{c}{Ring + Vortex} 		&& \multicolumn{1}{c}{Ring} 			& \multicolumn{1}{c}{Ring + Vortex}	\\
	 		& 					&			& \multicolumn{1}{c}{$\chi^2_{red}=1.18$}& \multicolumn{1}{c}{$\chi^2_{red}=1.05$}&& \multicolumn{1}{c}{$\chi^2_{red}=1.39$} 	& \multicolumn{1}{c}{$\chi^2_{red}=1.06$}	\\ 
\toprule
Ring			& $F_R$ 				& ($\mu$Jy)	& $5.72\pm0.01$		& $5.24_{-0.01}^{+0.03}$	&& $5.91\pm0.01$		& $4.72\pm0.02$		\\
Parameters	& $r_R$ 				& (AU)		& $59.9\pm0.1$		& $65.3\pm0.1$		&& $36.4\pm0.1$		& $35.0\pm0.1$		\\
			& $\sigma_R$ 			& (AU)		& $18.0\pm0.1$		& $15.5\pm0.1$		&& $14.90\pm0.04$		& $13.9\pm0.1$		\\
\cmidrule{1-8}
Vortex		& $F_V$ 				& ($\mu$Jy)	& -					& $7.3_{-0.3}^{+0.1}$	&& -					& $3.91_{-0.03}^{+0.02}$ \\
Parameters	& $r_V$ 				& (AU)		& -					& $41.5_{-0.1}^{+0.2}$	&& -					& $46.0_{-0.1}^{+0.2}$ 	\\
			& $\theta_V$ 			&($^\circ$)	& -					& $193.9_{-0.6}^{+0.4}$	&& -					& $177.7_{-0.3}^{+0.5}$	\\
			& $\sigma_{r,V}$		& (AU)		& - 					& $6.6_{-0.1}^{+0.3}$	&& -					& $14.4_{-0.1}^{+0.2}$	 \\
			& $\sigma_{\theta,V}$ 	& (AU)		& -					& $46.7_{-0.4}^{+0.7}$	&& -					& $40.4\pm0.4$		\\
\cmidrule{1-8}
Disk Geometry & $i$ 				& ($^\circ$)	& $22.9_{-0.2}^{+0.3}$	& $22.8_{-0.3}^{+0.2}$	&& $15.0_{-0.4}^{+0.3}$	& 15 (fixed) 			\\
			& $x_0$ 				& (mas)		& $-44.1\pm0.4$ 		& $-50.0_{-0.4}^{+0.6}$	&& $-12.5\pm0.4$		&$-17.6_{-0.6}^{+0.4}$	 \\
			& $y_0$ 				& (mas)		& $-51.6\pm0.5$		& $-75.0_{-0.2}^{+0.8}$	&& $-40.2\pm0.3$		& $2.7_{-0.7}^{+1.1}$	\\
\bottomrule
\end{tabular}
\end{center}
\end{table*}

\section{Observational Results}

We present maps of the dust  and  gas emission ($^{12}$CO~$J=6-5$) in Figure~1. 
These observations resolve the structure of each transitional disk, revealing for the first time the striking morphology in the dust continuum emission for \SAO\ and \SR:  a non-uniform ring with a bright asymmetry located in the south-west for \SAO\ and in the south for \SR. 
The integrated intensity map (moment 0) and the intensity-weighted mean velocity map (moment 1) of $^{12}$CO~$J=6-5$ are presented in the middle and right panels of Figure~1.
In contrast with the dust emission, the gaseous component from the $^{12}$CO moment 0 map appears to be quite compact and mostly symmetric.
However, significant emission ($>3\sigma$) can be found at large radii in the $^{12}$CO moment 1 map, just as with the dust emission. 
Since emission from $^{12}$CO is most likely optically thick, it is not a good tracer of depletion inside the cavity 
and other tracers must be used \citep{2013A&A...559A..46B}. However, a large depletion of the inner disk \citep[$>10^6$,][]{2013A&A...559A..46B} can give rise to the double-peaked structure observed in the \SAO\ moment 0 map. Further analysis of these CO observations will be deferred to a future publication.
We measured the disk position angle (PA) from the rotation of the disk in the  $^{12}$CO moment 1 map; the measured PA East of North is $\sim62^{\circ}$ and $\sim16^{\circ}$ for \SAO\ and \SR, respectively. 
Our derived PA is consistent with measurements found in the literature. 
For \SAO\ measurements from molecular gas (inside and outside the cavity) and polarization give a PA between $55^{\circ}$-$64^{\circ}$ \citep{2008ApJ...684.1323P,2011AJ....142..151L,2012Muto,2011ApJ...732...42A}. 
For \SR, \citet{2008ApJ...684.1323P} constrained its PA to $16^{\circ}\pm3^{\circ}$ from rovibrational CO band observations of molecular gas inside the cavity, while \citet{2009ApJ...704..496B} inferred a PA of $\sim 15^{\circ}$ from dust continuum observations of the outer disk.

The contrast between the peak of emission and the opposite side of the disk is $1.5\pm0.02$ in \SAO\ and $1.9\pm0.01$ in \SR. These asymmetries have lower contrast  values than those found in \LkHa\ \citep[$\times3$,][]{2013ApJ...775...30I}, \HD\ \citep[$\times30$,][]{2013Natur.493..191C,2013PASJ...65L..14F}, and \OphIRS\ \citep[$\times130$,]{2013Sci...340.1199V}, nevertheless they are significant enough to warrant further analysis.

The $850\:\mu$m observations from \citet{2011ApJ...732...42A} constrained the surface density at the edge of the cavity to be $\Sigma_{dust}\sim0.1\:\textrm{gm~cm}^{-2}$ for both \SAO\ and \SR. Assuming a dust opacity of $\kappa_{\nu}=10\:\textrm{cm}^{2}\:\textrm{g}^{-1}$  at $450\:\mu$m \citep{1990AJ.....99..924B}, we find that the 0.45~mm optical depth of dust is $\tau\sim1$ at the peak of the emission from the ring, and it is expected that at larger radii from the disk $\tau$ should be lower. 
We also computed the brightness temperature ($T_b$) of the emission without the Raleigh-Jeans approximation. The left panels of Figure~1 show contours of $T_b$, starting at 10~K and spaced every 5~K. At the peak of emission, $T_b\sim30$~K for \SAO, and $T_b\sim35$~K for \SR. At the distance where the asymmetry is located ($\sim50$~AU for \SAO\ and$\sim40$~AU for \SR), a temperature of 30-35~K is a reasonable value for the dust temperature in a flared disk, consistent with $\tau\approx1$ derived above.
We expect then that these ALMA observations  trace some of the mass surface density, since the emission is borderline between optically thin and optically thick emission.

\begin{figure*}[!t]
\begin{center}
\includegraphics[scale=0.17]{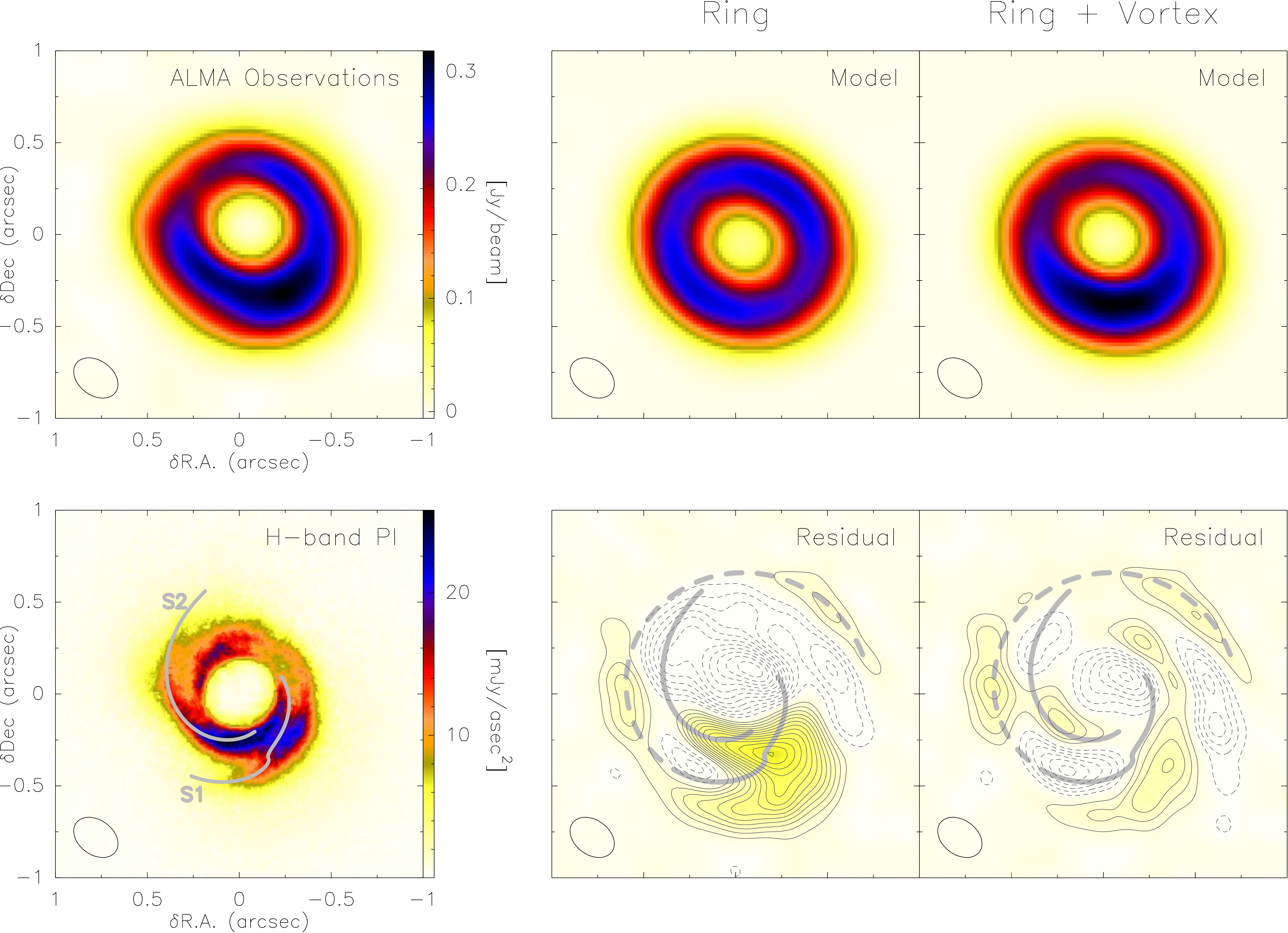}

\caption{\SAO\ ALMA observations (top left panel) compared with best-fit models for a ring (middle panels) and a ring with a vortex prescription (right panels). Residual maps were produced by subtracting the best-fit model from the \SAO\ ALMA observations. Contours start at $\pm3\sigma$, successively spaced by $\pm3\sigma$. Ellipses indicate beam size and colorbar defines scale used.
Bottom row, left panel: H-band polarized intensity image and the best-fit spiral arms from \citet{2012Muto}. Spirals S1 and S2 also shown with grey lines in residual maps.}
\label{SAO_model_fits}
\end{center}
\end{figure*}

\section{Analysis}

To constrain the structure of these large-scale asymmetries, we compare the dust continuum observations of \SAO\ and \SR\ with simple morphological models that describe the emission from a ring and a vortex. 
These morphological models were chosen since these are known transitional disks that can be well-described by a ring-like structure \citep{2011ApJ...732...42A,2009ApJ...704..496B}, and theoretical studies suggest that asymmetries in disk emission may arise for vortices \citep[e.g.,][]{2012MNRAS.419.1701R}. 
For the ring morphology we use a Gaussian prescription in the radial direction: in polar coordinates\footnote{where $r=\sqrt{x^2+y^2},\:\theta=\tan(x/y)$.} the ring emission corresponds to $F(r,\theta)=F_R(\theta)e^{-(r-r_R)^2/2\sigma_R^2}$, where $r_R$ is the radius where the ring peaks, $F_R$ is the flux density at $r_R$, and $\sigma_R$ is the ring width. In our modeling the ring emission is assumed to be axisymmetric. 
For the vortex morphology, we employ the prescription presented in \citet{2013Lyra}, where the mass-surface density of a steady-state vortex  can be described as a Gaussian in the radial and azimuthal directions: $F(r,\theta)=F_Ve^{-(r-r_V)^2/2\sigma_{r,V}^2}e^{-(\theta-\theta_V)^2/2\sigma_{\theta,V}^2}$, where $r_V$ and $\theta_V$ are the radius and PA of the vortex's peak, $F_V$ is the flux density at $r_V,\theta_V$, and $\sigma_{\theta,V},\sigma_{r,V}$ are the radial and azimuthal width of the vortex.
This way, the structure of a disk can be described by a ring of emission (3 parameters), a vortex (5 parameters), or the combination of both (8 parameters).

We fix the disk PA to the value measured in Section~3, and setup models of a ring and a ring with a vortex, leaving the disk center ($x_0,y_0$) and inclination ($i$) as free parameters. 
Each model realization is Fourier transformed and sampled at the same {\it uv}-positions as the observation. 
The best-fit model was found by minimizing the $\chi^2$ statistic using the latest implementation of the affine-invariant MCMC \emph{emcee} software \citep{2013PASP..125..306F}.

Table~2 presents the best-fit parameters found for a ring model and a ring+vortex model. 
As expected, the $\chi^2$ analysis favors a ring with a vortex model, since a ring alone cannot reproduce the large asymmetrical structures of these ALMA observations. 
We note that for SR 21, our preliminary ring+vortex model fitting favored a face-on geometry ($i\approx0^{\circ}$). 
Nevertheless, gas tracers show disk rotation making such geometry unlikely.
Thus, we fixed the inclination to $i=15^{\circ}$ for SR 21 ring+vortex modeling, 
consistent with the $^{12}$CO~$J=6-5$ line shape and motivated by the result 
of our ring-only model.

The best-fit model and residual maps for the two morphologies explored are presented in Figures 2 and 3 for \SAO\ and \SR, respectively.
These figures compare ALMA observations (top row, leftmost panel) with the best-fit model and residual maps (middle and right panels), where the residual emission is found by subtracting the model from the observations. When considering only a ring, the best-fit model corresponds to the average flux of the disk over all azimuthal angles. Hence large regions of negative emission are expected since the azimuthally averaged flux also encompasses flux from the large-scale asymmetry. The addition of a vortex prescription in the fitting improves the resulting residual map and $\chi^2$ of the fit, although significant emission ($\gtrsim12\sigma$) remains.

\begin{figure*}[!t]
\begin{center}
\includegraphics[scale=0.17]{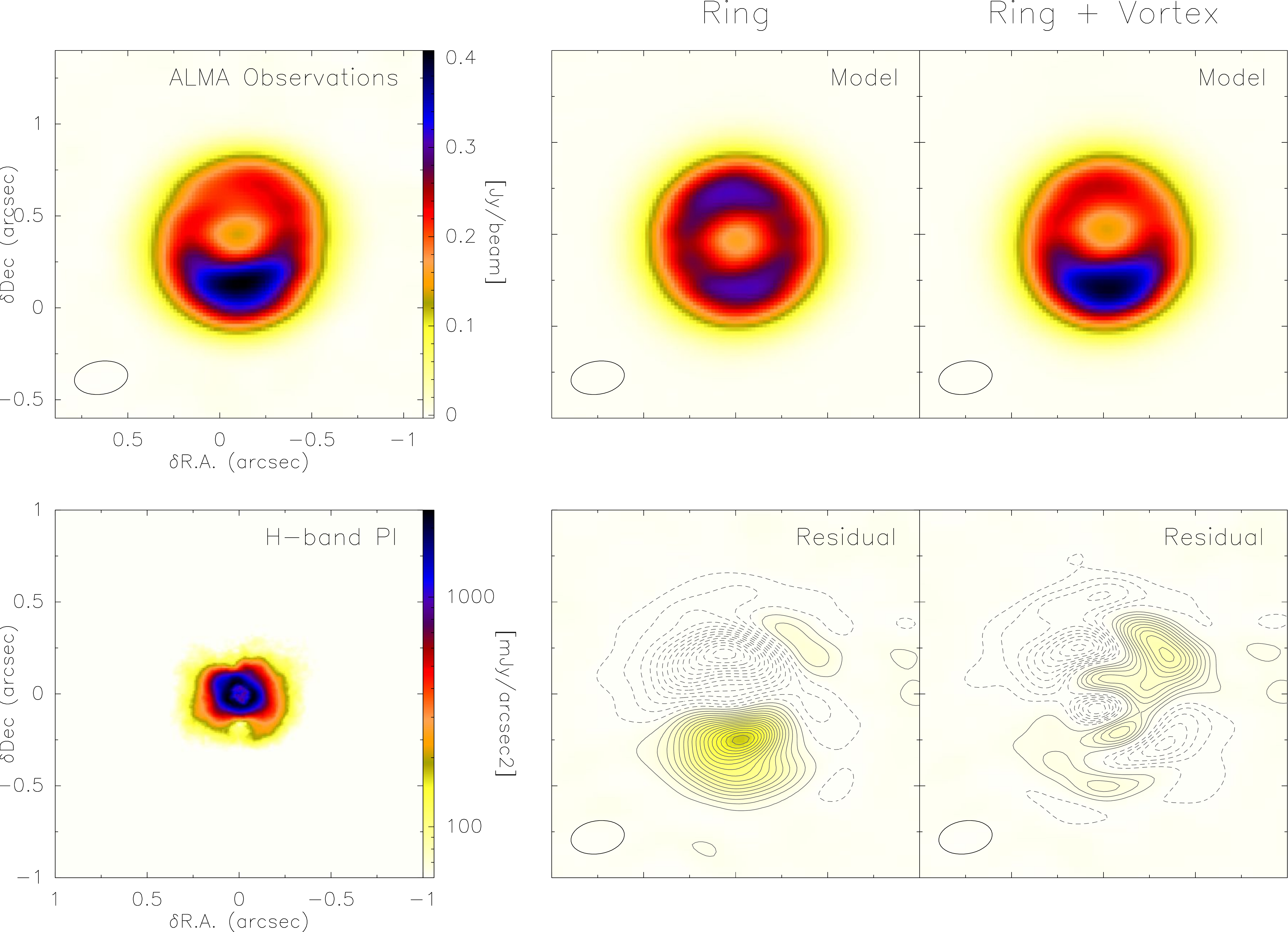}
\caption{\SR\ ALMA observations (top left panel) compared with best-fit models for a ring (middle panels) and a ring with a vortex prescription (right panels). Residual maps were produced by subtracting the best-fit model from the \SR\ ALMA observations. Contours start at $\pm3\sigma$, successively spaced by $\pm3\sigma$ ($\pm6\sigma$ for ring residual). Ellipses indicate beam size and colorbar defines scale used.
Bottom row, left panel: H-band polarized intensity image from \citet{2013ApJ...767...10F}.}
\label{SR_model_fits}
\end{center}
\end{figure*}

\section{Discussion}

The cavity radius ($R_{\rm cav}$) found in the literature for these disks has some scatter, 
on average $R_{\rm cav}\sim45$~AU for \SAO\ and $R_{\rm cav}\sim35$~AU for \SR\ 
\citep{2009ApJ...704..496B,2009ApJ...700.1502A,2011ApJ...732...42A,2011AJ....142..151L}.
We find that $R_{\rm cav}$ from these studies is consistent with our constraints within $r_R \pm \sigma_R$ 
for all but \citet{2009ApJ...704..496B} observations of \SAO, which have lower angular resolution than these ALMA observations.

The analysis presented above demonstrates that a ring of emission does not describe the disk structure of \SAO\ and \SR; 
the addition of a vortex prescription better describes the large-scale asymmetries. 
However, the observed asymmetries might not directly trace the mass surface density, since short-wavelength observations may not be completely optically thin. Future long-wavelength observations will confirm that these asymmetries correspond to mass rather than temperature variations.

The vortex encompasses 15\% of the total flux density for \SAO\ and 28\% for \SR. These are quite similar to the recently studied disk around \LkHa\ (30\%) but much less dramatic than the \OphIRS\ and \HD\ disks, where a horseshoe shape, rather than a ring, is observed at millimeter wavelengths.
The flux density in the asymmetry corresponds to a mass (dust+gas) of $\sim2\:M_{\textrm{Jup}}$ for both \SAO\ and \SR, assuming a temperature of $T=35K$, a gas-to-dust ratio of 100, and a dust opacity of $\kappa_{\nu}=10\:\textrm{cm}^{2}\:\textrm{g}^{-1}$ at $450\:\mu$m \citep{1990AJ.....99..924B}. These are lower limits on the vortex mass, given the uncertainty in  optical depth at this wavelength. 
Compared to recent examples of large-scale asymmetries observed in \LkHa\ and \OphIRS, the radial extent of the vortices in \SAO\ and \SR\ is also quite narrow, with a FWHM of few tens of AU (see Table~2). 
Azimuthally, the vortices are well resolved, and while the \SR\ asymmetry has a FWHM$\sim120^{\circ}$, similar to \LkHa\ and \OphIRS, the azimuthal extent in \SAO\ is 30\% larger, covering almost half of the disk.

From our modeling we find that the vortex aspect ratio is $\sigma_{\phi,V}/\sigma_{r,V}=7.1$ for \SAO\ and $\sigma_{\phi,V}/\sigma_{r,V}=2.8$ for \SR.
Following the solution presented by  \citet{2013Lyra}, a steady-state vortex tends to concentrate large grains toward its center, in an amount that depends on the ratio between the dust grain Stokes number ($St$) and the gas turbulence velocity (see also Birnstiel et al. 2012). 
Our measurements of the vortex structure and knowledge of the Stokes number allow us to constrain the gas turbulence.  
Dust grains with size $a=1$~mm and internal density $\rho_s=1$ g cm$^{-3}$, embedded in a disk with a surface density of $\Sigma=10$ g cm$^{-2}$ have a Stoke number $S_t\sim\rho_sa/ \Sigma\sim10^{-2}$.
The vortex aspect ratio derived for \SAO\ and \SR\ would then require turbulent velocities of about $90\:\textrm{m~s}^{-1}$ and $68\:\textrm{m~s}^{-1}$ respectively, corresponding to about 22\%  and 16\% of the local isothermal sound speed for a gas 
temperature of 50~K. Such values are about 5 times larger that those derived for \OphIRS, and consistent with an active disk layer due to the magnetorotational instability \citep{2011ApJ...743...17S}.

When including the vortex prescription in the fit, there are still significant residual emission in both \SAO\ and \SR. These residuals are not symmetric, and their morphology is not well described by an additional vortex since they extends outwards with increasing radius.
A spiral-like structure may be a good approximation for the residual emission seen in both disks, but the number of parameters of such a prescription is quite large (at least 5 more parameters per spiral arm). Given the low signal of these features and degeneracies between such large number of parameters, we decided not to proceed with such a fit. Observations at higher angular resolution are required to study in greater detail the morphology of these structures.

On the bottom left panel of Figures 2 and 3, we present polarized intensity maps at H-band of both disks \citep{2012Muto,2013ApJ...767...10F}. This emission, likely optically thick, does not directly traces mass surface density.
In \SAO\ two spiral-like structures are seen in H-band imaging (Figure 2). 
\citet{2012Muto} compared the shape of these structures with theoretical models for a perturbation due to low-mass companions, their best-fit spirals, S1 and S2, are labeled in Figure 2.
For the S1 spiral, the location of the unseen perturber coincides with the location of the peak of emission in the ALMA map (at $\sim55$~AU from the star with PA of $204^{\circ}$). 
If the 0.45~mm observations trace a mass overdensity at this location, we may be witnessing a spiral wave pattern that arose from a density enhancement rather than a point-like perturber, as reviewed by \citet{2010ApJ...725..146P}.
Figure 2 also shows the location of S1/S2 with respect to the modeling residuals. 
Some overlap exist between the submillimeter residuals and the near-IR emission, with the S1/S2 spiral arms generally found inward of where the submillimeter residuals are. Nevertheless, observations with better angular resolution of \SAO\ are required to investigate whether the submillimeter emission traces the spiral structures observed in the near-IR.

In the case of \SR, \citet{2013ApJ...767...10F} showed that not only is the H-band polarized intensity image is rather smooth and axisymmetric, but it is also not compatible with the large depletion of material in the inner disk ($10^{-6}$) found by \citet{2011ApJ...732...42A}. 
Furthermore, radiative transfer modeling of these near-IR observations did not find a transition between the outer disk and the depleted inner disk, even for modest depletion levels of 1-10\% \citep{2013ApJ...767...10F}. 
These H-band observations of \SR\ show evidence for emission from small grains out to a 80~AU radius, and compared with our ALMA observations, it shows no signs of the observed asymmetries (Figure~3).
The difference in spatial distribution of the small grains (traced by near-IR observations) and the mm-sized particles (traced by these ALMA observations), could arise from the presence of a local enhancement in the gas density. 
For example disk vortices, which can be excited by Rossby-wave instabilities at the outer edge of gas depleted cavities, are particularly efficient at trapping large dust grains while small grains remain coupled to the gas \citep{2013A&A...550L...8B}.
Although not as dramatic as in the \OphIRS\ case, the vortex in \SR\ may be driving the distribution of large and small particles within the disk.

\section{Conclusions}

We have observed two transitional disks, \SAO\ and \SR, with high sensitivity using ALMA. 
These transitional disks display large-scale asymmetries in their dust continuum emission, just as two other recently observed with ALMA \citep{2013Natur.493..191C,2013PASJ...65L..14F,2013Sci...340.1199V}, and an additional disk observed with CARMA \citep{2013ApJ...775...30I}.
It is interesting to note that of the few such objects observed with high sensitivity and high angular resolution, most exhibit high contrast asymmetrical structures. 
The lack of previous detections is likely due to phase errors introduced by the atmosphere and the lack of sensitivity of previous observations. 
This may imply that the process (or processes) that drive these asymmetries is taking place in most transitional disks.

In the case of \SAO\ and \SR, we have demonstrated that a ring does not describe these observations, and a vortex-like asymmetry, in addition to a ring of emission, is a better match. Such asymmetrical structures encompass a significant fraction ($\sim20\%$) of the total disk emission, although we found that their contrast is much smaller than in \LkHa, \HD, and \OphIRS. Asymmetries in  \SAO\ and \SR\ have a narrow radial size and a broad azimuthal extent, similar to the \LkHa\ and \OphIRS\ asymmetrical structures. If these asymmetries are in fact non-transient vortices, the turbulence inside them needs to be substantial in order to produce the azimuthally-broad asymmetrical features. 

Despite being a better fit, the vortex prescription does not reproduce every observed feature. 
Significant residuals remain that are suggestive of spiral-like structure, in the case of 
\SAO\ some residuals roughly coincide with the spiral arms seen in H-band scattered light. 
To test the presence of large-scale spiral-like features, and to make sure the observed structures 
trace the mass surface density of the disk down to the midplane, observations at longer wavelengths 
are required. Future observations with ALMA at 3~mm, and/or with the Very Large Array at 7~mm, will 
help in this regard, since such observations trace larger dust grains and will have lower optical depth.

\acknowledgments
We thank Takayuki Muto, Carol Grady, and Katherine Follette for providing the near-IR images, and we thank Wladimir Lyra for useful discussions.
A.I., J.M.C., L.M.P acknowledge support from NSF award AST-1109334.
The National Radio Astronomy Observatory is a facility of the National Science Foundation operated under cooperative agreement by Associated Universities, Inc. This paper makes use of the following ALMA data: ADS/JAO.ALMA 2011.0.00724.S. ALMA is a partnership of ESO (representing its member states), NSF (USA) and NINS (Japan), together with NRC (Canada) and NSC and ASIAA (Taiwan), in cooperation with the Republic of Chile. The Joint ALMA Observatory is operated by ESO, AUI/NRAO and NAOJ.

{\it Facilities:}\facility{ALMA}.

\end{document}